\documentclass[12pt]{article}
\usepackage[symbol*]{footmisc}
\usepackage{array}
\usepackage{amsmath}
\usepackage{amssymb}
\usepackage{amsfonts}
\usepackage{graphicx}

\def\la{~\mbox{\raisebox{-.6ex}{$\stackrel{<}{\sim}$}}~}
\def\ga{~\mbox{\raisebox{-.6ex}{$\stackrel{>}{\sim}$}}~}

\begin{document}

\thispagestyle{empty}
\vspace*{1cm}
\begin{center}
{\Large \bf {Non-Gaussianities and chiral gravitational waves}}\\ 
\vspace*{0.3cm}
{\Large \bf {in natural steep inflation}}\\
\vspace*{1.5cm} {\large Mohamed M. Anber$^{a,}$\footnote{\tt
manber@physics.utoronto.ca} and
Lorenzo Sorbo$^{b,}$\footnote{\tt sorbo@physics.umass.edu}}\\
\vspace{.5cm}  
{\em $^{a}$ Department of Physics,
University of Toronto, Toronto, ON, M5S1A7, Canada}\\
\vspace{.15cm} {\em $^{b}$ Department of Physics,
University of Massachusetts Amherst, MA 01003, USA}\\
\vspace{1.5cm}

ABSTRACT

\end{center}

{In~\cite{Anber:2009ua}, we have proposed a model where natural inflation is realized on a steep potential ($V(\phi)\sim \cos(\phi/f)$ with $f\ll M_P$) as a consequence of the interaction of the inflaton with gauge fields through the coupling $\phi\,F_{\mu\nu}\,\tilde{F}^{\mu\nu}$. In the present work we study the nongaussianities and the spectrum of tensor modes generated in this scenario. The nongaussianities turn out to be compatible with current observations and can be large enough to be detectable by Planck. The non-observation of tensor modes imposes new constraints on the parameter space of the system that are about one order of magnitude stronger than those found in~\cite{Anber:2009ua}. More importantly, in certain regions of the parameter space tensor modes might be detected by upcoming Cosmic Microwave Background experiments even if inflation occurs at energies as low as the TeV scale. In this case the tensor modes would be chiral, and would lead to distinctive parity-violating correlation functions in the CMB.
}

\vfill \setcounter{page}{0} \setcounter{footnote}{0}

\section{Introduction}%

An axion-like field $\Phi$ with 
\begin{equation}\label{cos}
V(\Phi)=\frac{\Lambda^4}{2}\left[\cos\left(\frac{\Phi}{f}\right)+1\right]
\end{equation}
represents the simplest example of spin-0 degree of freedom with a nontrivial and yet  radiatively stable potential. The technically natural flatness of $V(\Phi)$ makes $\Phi$ an ideal inflaton candidate~\cite{Freese:1990rb}. In all known UV-complete models including potentials of the form~(\ref{cos}), the parameter $f$ is significantly smaller than the reduced Planck mass $M_P=1/\sqrt{8\pi G}$~\cite{Banks:2003sx}. It is actually conjectured~\cite{ArkaniHamed:2006dz} that it is impossible for a UV-complete theory that includes gravity to fulfill the condition $f>M_P$: the potential~(\ref{cos}), while {\em controllably} flat, cannot be {\em arbitrarily} flat. Unfortunately, the condition $f>M_P$ is precisely what~(\ref{cos}) requires to support inflation.

There exists by now a substantial body of literature (see e.g.~\cite{Kim:2004rp,Dimopoulos:2005ac,McAllister:2008hb,Kaloper:2008fb,Kaloper:2011jz}) that aims at maintaining the properties of radiative stability of the axion-like field while changing the shape of $V(\Phi)$ to allow it to support inflation. In~\cite{Anber:2009ua}, we have taken a different approach. Instead of deforming $V(\Phi)$, we have coupled $\Phi$ to matter. More specifically, we have considered the effect on the dynamics of $\Phi$ induced by its coupling to a $U(1)$ gauge field $A_\mu$ via
\begin{equation}\label{coupl}
{\cal L}_{\Phi F\tilde{F}}=\frac{\alpha}{f}\,\Phi\,F_{\mu\nu}\,\tilde{F}^{\mu\nu}\,\,,
\end{equation}
with $\alpha$ a dimensionless parameter. In the presence of this coupling, the rolling inflaton uses a finite fraction of its kinetic energy to produce quanta of $A_\mu$. This coupling has thus the effect of slowing down the rolling of $\Phi$, making it possible to achieve inflation even if $V(\Phi)$ is steep (i.e., $f\ll M_P$).  This situation is analogous to that leading to warm inflation~\cite{astro-ph/9509049,arXiv:0808.1855,Visinelli:2011jy}, to trapped inflation~\cite{Green:2009ds} and, more recently, to the mechanism discussed in~\cite{Dimopoulos:2010xq}. We also note that a model related the present scenario has been recently proposed in~\cite{Adshead:2012kp,Adshead:2012qe}, where the inflaton is coupled to $SU(2)$ gauge fields leading to a classical, rotationally invariant configuration able to support slow-roll even for $f<M_P$.

The analysis of~\cite{Anber:2009ua} shows that it is possible to obtain sufficiently long inflation, irrespective of the value of $f$, if $\alpha$ is of the order of a few hundred. In the simplest version of the model the amplitude of the scalar perturbations is too large to match observations. This happens (almost) irrespective of the parameters in the theory because, once $\alpha$ is large enough to allow for sufficient inflation, the observables of this model have a very weak, logarithmic dependence on the parameters of the theory (see section 2).  It is however possible to reduce the amplitude of the scalar perturbations to its observed value by considering a modification of the theory where the inflaton couples to ${\cal N}\simeq 10^5$ $U(1)$ gauge fields.

Is this model in any sense generic? Of course it is not -- both the large value of $\alpha$ and the large number of gauge fields are not generic predictions of UV-complete theories. However, no fully working model of inflation is generic. And actually, in the spirit of chaotic inflation, inflation does not {\em need} to be generic: as long as some corner of the parameter space of the theory allows it, inflation will happen and the inflating part of the multiverse will overwhelm the remaining sections. In~\cite{Anber:2009ua} we have discussed situations where the conditions $\alpha={\cal {O}}(10^2)$ and ${\cal {N}}={\cal {O}}(10^5)$ can be consistently realized. In this sense, this set-up is as generic as we can require.

The only way of discriminating between models is to study their predictions and compare them to observations. In~\cite{Anber:2009ua} we have shown that the spectral index of scalar perturbations is $n\simeq 1+(2/\pi\alpha)\,(f\,V''/V')$, and so it is extremely close to $1$. More recently, \cite{Barnaby:2010vf,Barnaby:2011vw} have studied the nongaussianities and~\cite{Sorbo:2011rz} has studied the spectrum of (chiral) gravitational waves induced by the coupling~(\ref{coupl}), that might even be observable by ground-based gravitational interferometers in the near future~\cite{Cook:2011hg}. The analyses of refs.~\cite{Barnaby:2010vf,Barnaby:2011vw,Sorbo:2011rz,Cook:2011hg} (see also~\cite{Barnaby:2011qe}) are however performed in the regime where the backreaction of the gauge field on the rolling inflaton is negligible, and therefore inflation is obtained because $V(\Phi)$ (that is otherwise considered to have an arbitrary shape) is sufficiently flat.

In the present paper we study the nongaussianities and the spectrum of tensor modes in the model~\cite{Anber:2009ua}. We find that the $f_{NL}$ computed on equilateral configurations takes values in the range $f_{NL}^{\rm {equil}}\sim -30\div -5$ depending on the energy scale of inflation. Nongaussianities are therefore within the current bounds, and could be detected by Planck. 

Our predictions for the spectrum of tensor modes depend on the value of the coupling $\alpha$ and on the scale of inflation. We find that non-observation of tensor modes induces a bound $\alpha\ga 300\div 10^4$ depending on the scale of inflation. Moreover, as usual, the scale of inflation has to be below the $10^{16}$ GeV if we want to avoid overproduction of tensors. Remarkably, however, the tensors-to-scalar ratio can be ${\cal O}(10^{-1})$, and therefore detectable by upcoming CMB polarization experiments, irrespective of the energy scale at which inflation occurs, provided $\alpha$ has the correct magnitude. The system of~\cite{Anber:2009ua} provides therefore an example of a model where gravitational waves of inflationary origin could be detected even if inflation happens at energies much smaller than the GUT scale~\cite{Senatore:2011sp}. Unlike tensors produced by most other mechanisms, these gravitational waves would be fully chiral~\cite{Sorbo:2011rz}, a property that would show up in parity-violating correlation functions in the CMB, providing a very specific signature of this scenario.

The paper is organized as follows. In section 2 we review the model~\cite{Anber:2009ua} and study the evolution of the homogeneous component of the inflaton. In section 3 we provide general formulae that allow to compute the spectra of scalar perturbations in this system. In section 4 we give explicit expressions for the scalar power spectrum and bispectrum in terms of the main parameters of the theory. In section 5 we compute the spectrum of tensor modes and in section 6 we discuss our results.

\section{Natural steep inflation}%

In order to make the paper self-contained, in this section we review the model and the results of~\cite{Anber:2009ua}. We consider natural inflation with a pseudoscalar inflaton $\Phi$ coupled to ${\cal N}$ $U(1)$ gauge fields. The Lagrangian density of the system is
\begin{eqnarray}
{\cal L}=-\frac{1}{2}(\partial\Phi)^2-V(\Phi)-\frac{1}{4}F^a_{\mu\nu}F_a^{\mu\nu}-\frac{\alpha}{4\,f}\Phi F^a_{\mu\nu}\tilde F_a^{\mu\nu}\,,
\end{eqnarray}
where $V(\Phi)=\Lambda^4\left[1+\cos(\Phi/f) \right]$ with $f \la M_P$. The index $a$ is understood to be summed upon, and it labels the $U(1)$ groups, ranging from $1$ to ${\cal N}$. The parameter $\alpha$ is the dimensionless coupling constant between axion and gauge field. Introducing the vector potential ${\bf A}^a(\tau,{\bf x})$ with $a^2\,{\bf B}^a=\nabla\times {\bf A}^a$, $a^2\,{\bf E}^a=-{\bf A}^a{}'$, we obtain the equations of motion
\begin{eqnarray}
\nonumber
\Phi''+2\,a\,H\Phi'-\nabla^2 \Phi+a^2\frac{dV(\Phi)}{d\Phi}=\frac{\alpha}{f}\,a^2\,{\bf E}^a \cdot {\bf B}^a\,,\\
\label{equations of motion}
\left(\frac{\partial^2}{\partial \tau^2}-\nabla^2-\alpha\frac{\Phi'}{f}\,\nabla \times  \right){\bf A}^a=0\,,\quad \nabla\cdot {\bf A}^a=0\,,
\end{eqnarray}
where $H=a'(\tau)/a^2(\tau)$ and where the prime denotes differentiation with respect to the conformal time $\tau$.

\subsection{Generation of the gauge field}%

The rolling inflaton induces the generation of quanta of the gauge field. In order to study this process we promote the classical field ${\bf A}^a(\tau,\,{\bf x})$ to an operator $\hat{\bf A}^a(\tau,\,{\bf x})$. We decompose $\hat{\bf A}^a$ into annihilation and creation operators 
\begin{eqnarray}
{\bf A}^a(\tau,\,{\bf x})=\sum_{\lambda=\pm}\int \frac{d^3 k}{\left(2\pi\right)^{3/2}}\left[{\bf e}_\lambda({\bf k})\,A^a_{\lambda}(\tau,\,{\bf k})\,\hat a_\lambda({\bf k})\,e^{i\,{\bf k} \cdot {\bf x}} +{\mbox {h.c.}} \right]\,,
\end{eqnarray}
where the helicity vectors ${\bf e}_\pm({\bf k})$ are defined in such a way that ${\bf k} \cdot {\bf e}_{\pm}=0$, ${\bf k} \times {\bf e}_{\pm}=\mp i\,k\,{\bf e}_{\pm}$. Then, the mode functions $A^a_{\pm}$ must satisfy the equations $A^a_{\pm}{}''+\left(k^2\mp\alpha\,k\,\Phi'/f\right)\,A^a_{\pm}=0$. 

Since we are looking for inflating solutions, we assume $a(\tau)\cong -1/(H\tau)$, and $d\Phi/dt\equiv \dot \Phi_0=\mbox{constant}$. Hence, the equation for $A^a_\pm$ reads
\begin{eqnarray}\label{eqapm}
\frac{d^2 A^a_\pm (\tau,k)}{d\tau^2}+\left[k^2\pm 2\,k\,\frac{\xi}{\tau} \right]\,A^a_\pm (\tau,k)=0\,,
\end{eqnarray}
where we have defined 
\begin{eqnarray}
\xi\equiv\alpha \frac{\dot \Phi_0}{2\,f\,H}\,,
\end{eqnarray}
and we will be interested in the case $\xi \ga {\cal O}(1)$. Depending on the sign of $\xi$, one of the two solutions $A^a_+$ or $A^a_-$ of eq.~(\ref{eqapm}) will develop an instability. We assume that  $\alpha>0$ and $V'(\Phi)<0$,  so that $\xi>0$.

The solution that reduces to positive frequency for $k\,\tau \to -\infty$ is
\begin{eqnarray} 
 A^a_\pm (\tau,\,k)=\frac{1}{\sqrt{2\,k}}\left[i\,F_0\left(\pm \xi,-k\,\tau \right)+G_0\left(\pm \xi,-k\,\tau \right) \right]\,,
\end{eqnarray} 
where $F_0$ and $G_0$ are the regular and irregular Coulomb wave functions respectively. At late times $|k\,\tau|\ll 2\,\xi$ the positive helicity mode $A_+$ behaves as 
\begin{eqnarray}\label{approximate solution}
A^a_+(\tau, \vec k)\cong\frac{1}{\sqrt{2\,k}}\left(\frac{k}{2\,\xi\,aH}\right)^{1/4}e^{\pi\xi-2\sqrt{2\,\xi\,k/aH}}\,,
\end{eqnarray}
and is thus amplified by a factor $e^{\pi \xi}$. On the other hand, the mode $A_-^a$ is not amplified by the rolling inflaton and we will ignore it from now on.

\subsection{The slow roll solution}%

The amplified gauge modes backreact on the rolling inflaton, and slow down its rolling along $V(\Phi)$. Here we describe the portion of parameter space where such effect is so large to lead to slow roll inflation on a steep potential. In order to estimate the backreaction of the gauge field on the inflaton, we use the decomposition of ${\bf A}^a(\tau,\,{\bf x})$ described above. Then, the quantity $\langle {\bf E}^a\cdot {\bf B}^a\rangle$ is very well approximated, for $\xi\ga 3$ (that, as we will see, is the regime we are interested in), by
\begin{eqnarray}
\langle {\bf E}^a\cdot {\bf B}^a\rangle\cong -2.4\times 10^{-4}\,{\cal N}\,\left(\frac{H}{\xi}\right)^4 e^{2\pi\,\xi}\,,
\end{eqnarray}
that we plug into the first of eqs.~(\ref{equations of motion}), to obtain the equation of motion of $\Phi$ (now written in the physical time $t$)
\begin{eqnarray}\label{equation of motion of Phi}
\frac{d^2 \Phi}{dt^2}+3\,H\,\frac{d\Phi}{dt}+V'(\Phi)=-2.4\times 10^{-4}\,{\cal N}\,\frac{\alpha}{f}\left(\frac{H}{\xi} \right)^4 e^{2\pi\xi}\,.
\end{eqnarray}
Since we are interested in finding inflationary solutions where slow roll is supported by the dissipation into electromagnetic modes, we assume that both $\ddot \Phi$ and $3H\dot \Phi$ are negligible with respect to $V'(\Phi)$. In this case, the expression of $\xi$ is controlled by the transcendental equation
\begin{equation}\label{the main expression for xi}
\xi=\frac{1}{2\pi}\log \left[\frac{1.3\times 10^4}{{\cal N}\,\alpha}\,\xi^4\,\frac{M_P^2}{H^2}\frac{f|V'(\Phi)|}{V(\Phi)} \right]
\end{equation}
where we have assumed $3\,M_P^2\,H^2=\frac{1}{2}\,\dot\Phi^2+V(\Phi)+\frac{1}{2}\,\left({\bf E}_a^2+{\bf B}_a^2 \right)\cong V(\Phi)$.

As we will see below, ${\cal {N}}={\cal O}(10^5)$, $\alpha={\cal O}(10^3)$ and $\xi={\cal O}(10)$, so that, using $V'\simeq V/f$, $\xi\simeq \frac{2}{\pi}\log \frac{M_P}{\Lambda}$.

As shown in~\cite{Anber:2009ua}, the slow roll conditions $\frac{1}{2}\dot \Phi^2+\frac{1}{2}\left({\bf E}_a^2+{\bf B}_a^2 \right)\ll V(\Phi)$, $|\dot{H}|\ll H^2$, $|\ddot\Phi+3\,H\dot\Phi|\ll |V'(\Phi)|$ can all be satisfied if
\begin{equation}\label{condend}
\frac{|f\,V'(\Phi)|}{V(\Phi)}\ll \frac{\alpha}{\xi}\,.
\end{equation}
When this condition is violated, the energy in the electromagnetic field cannot be neglected with respect to the energy in the inflaton. This happens when we are close to the bottom of the potential. Indeed, by approximating $V(\Phi)\propto \Phi^2$ near its minimum, we see that the condition~(\ref{condend}) is violated when $\Phi\sim \xi\,f/\alpha$. This point marks the beginning of reheating, so that the reheating temperature is estimated to be $T_{\rm RH}\simeq \Lambda\,\sqrt{\xi/\alpha}$. 

One more condition on the parameter space of the system comes from the requirement to have enough inflation. The number of efoldings is given by 
\begin{equation}\label{3rd constraints}
N_e\simeq\int_{\Phi_i}^{\Phi_f}\frac{H\,d\Phi}{\dot{\Phi}}=\frac{\alpha}{2f}\int_{\Phi_i}^{\Phi_f}\frac{d\Phi}{\xi}\simeq \frac{\alpha}{2\,\xi}\,\frac{\Phi_f-\Phi_i}{f}\,.
\end{equation}
Since the range of variation of $\Phi$ is bounded by $|\Phi_f-\Phi_i|\la\pi\,f$, the above equation implies that $\alpha/\xi\ga 2\,N_e/\pi$. 

How large can $\xi$ be? Let us say that we require the reheating temperature to be larger than ${\cal O}(10^2)$~GeV, so to leave room for some mechanism of baryogenesis related to the electroweak phase transition. Then since $\xi/\alpha\sim 10^{-2}$, we can take $\Lambda$ as low as a few TeVs, corresponding to $\xi\simeq 20$. In this case, since inflation occurs at a low energy scale, only $\sim 30$ efoldings of inflation are needed to solve the problems of standard Big Bang cosmology, and we will need $\alpha\ga 400$. In the opposite regime of very high energy inflation, $\xi$ can be as small as $4$ for $\Lambda\simeq 10^{16}$~GeV -- in which case $N_e\simeq 60$ so that we need $\alpha\ga 150$. As we will see in section 5, the requirement that the system does not lead to overproduction of tensors will impose stronger constraints on $\alpha$.

A numerical study of eq.~(\ref{equation of motion of Phi}) supports the validity of the analytical estimates above. We show in figure 1 the evolution of the scale factor, of $\xi$ and of the slow-roll parameter $\epsilon=-\dot{H}/H^2$ for the choice of parameters $\alpha=300$, $\Lambda=10^{-3}\,M_P$ and $f=0.1\,M_P$. The left panel of figure 1 shows that the parameter $\xi$, rather than being constant, increases with time. The time dependence of $\xi$ is however rather mild -- it starts at $\xi\simeq 4$ and ends at $\xi\simeq 6$ after about $60$ efoldings of inflation -- and is in approximate agreement with the rough estimate $\xi\simeq \frac{2}{\pi}\,\log \frac{M_P}{\Lambda}$ that would give $\xi\simeq 4.4$ for this choice of parameters.

\begin{figure}
\centering
    \includegraphics[width=7.5cm]{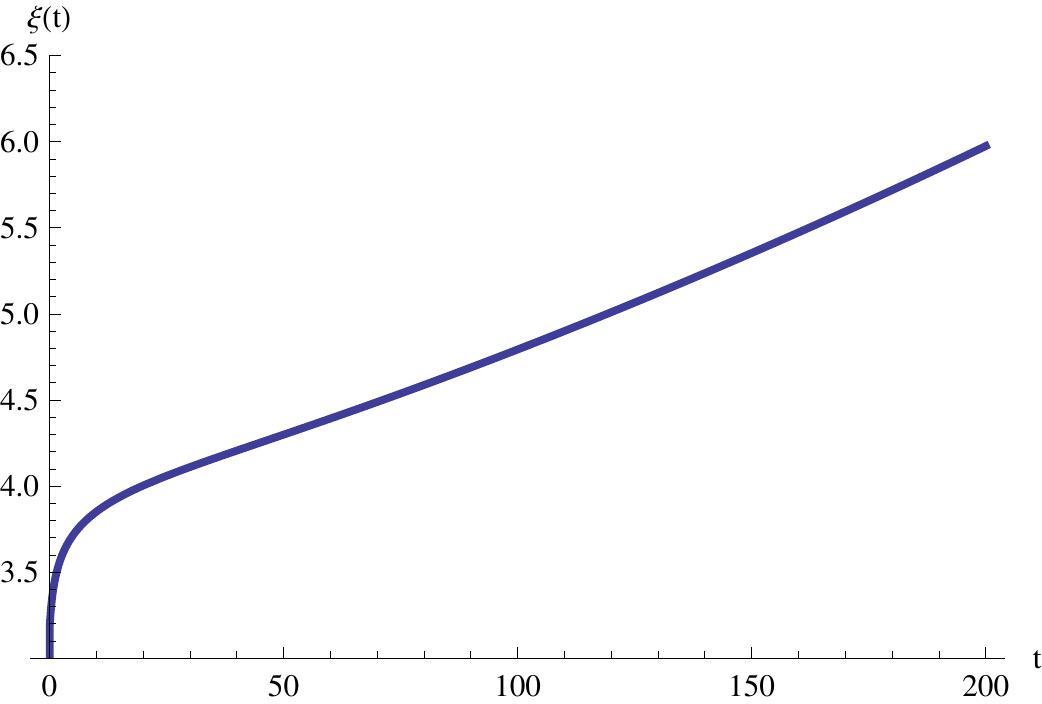}
    \includegraphics[width=7.5cm]{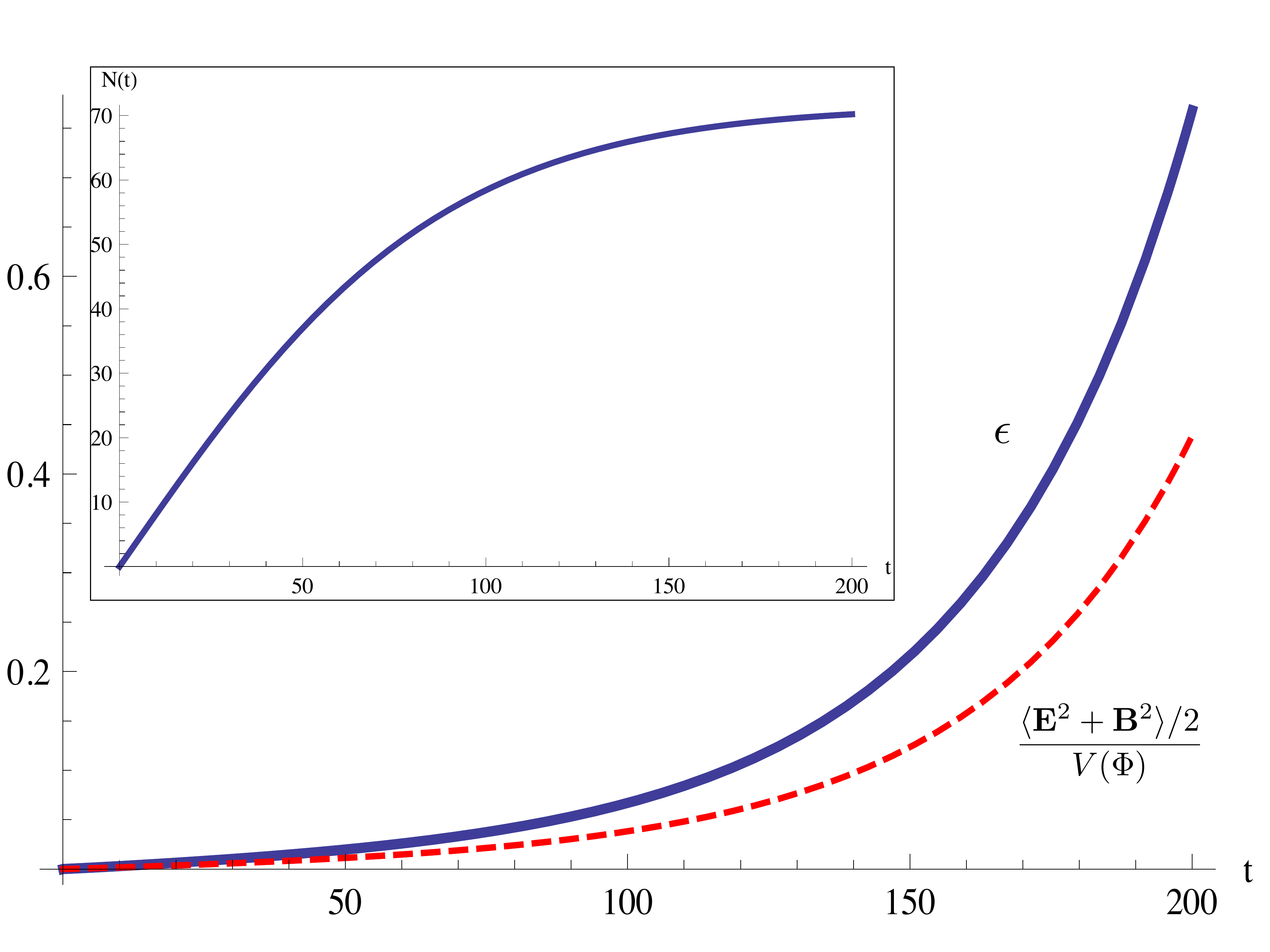}
\caption{Evolution of background quantities for $\Lambda=10^{-3}\,M_P$, $f=0.1\,M_P$, $\alpha=300$ and ${\cal N}=10^5$. Left panel: evolution of $\xi(t)$. Right panel: the solid line correspond to the slow-roll parameter $\epsilon=-\dot{H}/H^2$, the dashed line gives the ratio of the energy in gauge modes over the energy in the inflaton. In the inset we plot the number of efoldings as a function of time. The time $t$ is in units of $M_P/\Lambda^2$.}
\end{figure}

\section{Perturbations}%

Perturbations are usually generated as the quantum fluctuations of the inflaton are amplified by the evolving background. This is different from our scenario, where inhomogeneities in $\Phi$ are sourced by those in the electromagnetic field, analogously to the situations encountered in warm~\cite{astro-ph/9509049,arXiv:0808.1855} or in trapped inflation~\cite{Green:2009ds,Lee:2011fj}. 
 
The curvature perturbation $\zeta$ on a uniform energy density hypersurface is related to the perturbation of the number of efoldings by $\zeta=\delta _N\equiv N(x)-\tilde N$, where $\tilde N$ is the number of efoldings in the homogeneous background. If we write the perturbed value of the inflaton field as $\Phi=\Phi_0(\tau)+\phi(\tau,\vec x)$, then $\zeta =H\,\phi/\dot \Phi_0$. The perturbation $\phi$ obeys the equation
\begin{eqnarray}
\phi''+2\,a\,H\,\phi'+\left(-\nabla^2+a^2\,V''(\Phi_0)\right)\,\phi=-\frac{\alpha}{f}\,\delta[{\bf E}^a\cdot{\bf B}^a]\,,
\end{eqnarray}
where the fluctuation $\delta[{\bf E}^a\cdot{\bf B}^a]$ receives two contributions, the first  from the intrinsic inhomogeneities in  ${\bf E}^a\cdot {\bf B}^a$ (that would be present for $\phi=0$), and the second from the fact that $\left\langle {\bf E}^a\cdot {\bf B}^a \right\rangle$ depends on $\dot\Phi_0$
\begin{eqnarray}\label{deltaeb}
\delta[{\bf E}^a\cdot {\bf B}^a]=\left[{\bf E}^a\cdot {\bf B}^a-\left\langle {\bf E}^a\cdot {\bf B}^a \right\rangle  \right]_{\phi=0}+\frac{\partial \left\langle {\bf E}^a\cdot {\bf B}^a \right\rangle  }{\partial \dot\Phi_0}\dot \phi\,.
\end{eqnarray}
The above equation should also include a term proportional to $\phi$ through the dependence of ${\bf E}^a\cdot {\bf B}^a$ on $H$, but this contribution turns out to be subdominant.

We denote the term in brackets in eq.~(\ref{deltaeb}) by $\delta_{{\bf E}\cdot {\bf B}}$, and the second term gives $\pi\,\alpha\,V'\dot\phi/(f\,H)$ upon using the background equation $\alpha \left\langle {\bf E}^a\cdot {\bf B}^a\right\rangle/f \cong V'$. The  Fourier transform  of the perturbation $\phi$ will then obey the equation
\begin{eqnarray}\label{largeeqphi}
\phi''({\bf k})-\frac{2}{\tau}\left(1-\frac{\pi\,\alpha\,V'}{2\,f\,H^2}\right)\phi'({\bf k})+\left(k^2+\frac{V''}{H^2\,\tau^2}\right)\phi({\bf k})=-\frac{\alpha}{f}a^2\int \frac{d^3{\bf x}}{\left(2\pi\right)^{3/2}}\,e^{-i{\bf k}\cdot {\bf x}}\,\delta_{{\bf E}\cdot {\bf B}}(\tau,\,{\bf x})\,.
\end{eqnarray}
For steep potential inflation we have $\pi\,\alpha\,V'/(f\,H^2)\sim \alpha\,M_P^2/f^2>>1$. Moreover, we are interested in the super-horizon wavelengths $k<<a\,\sqrt{V''}$. Hence, the solution of this equation can be written as
\begin{eqnarray}
\phi({\bf k},\,\tau)=-\frac{\alpha}{f}\int_{-\infty}^\tau d\tau_1\,G(\tau,\,\tau_1)\,a^2(\tau_1)\,J(\tau_1,\,{\bf k})\,,
\end{eqnarray}
where $J(\tau,\,{\bf k})=\int \frac{d^3{\bf x}}{\left(2\pi\right)^{3/2}}\,e^{-i{\bf k}\cdot {\bf x}}\,\delta_{{\bf E}\cdot {\bf B}}(\tau,\,{\bf x})\,$, and the retarded Green's function $G(\tau,\,\tau')$ for the operator 
\begin{equation}
\frac{\partial\,\,\,}{\partial\tau^2}-\frac{1}{\tau}\frac{\pi\,\alpha\,V'(\Phi_0)}{f\,H^2}\,\frac{\partial\,\,}{\partial\tau}+\frac{V''(\Phi_0)}{H^2\,\tau^2}
\end{equation}
reads
\begin{eqnarray}
&&G(\tau,\,\tau')=\frac{\tau'}{\nu_+-\nu_-}\left[\left(\frac{\tau}{\tau'}\right)^{\nu_+}-\left(\frac{\tau}{\tau'}\right)^{\nu_-}\right]\Theta(\tau-\tau')\,,\\
&&\nu_\pm\equiv\frac{\pi\,\alpha\,V'}{2\,f\,H^2}\left(1\pm\sqrt{1-\frac{4}{\pi^2}\,\frac{V''\,H^2\,f^2}{\alpha^2\,V'^2}}\right).\nonumber
\label{nupm}
\end{eqnarray}
The second term under the square root in $\nu_\pm$ scales as $(f/\alpha\,M_p)^2$ and is much smaller than unity. As consequence
\begin{eqnarray}
&&\nu_+ \cong\pi\,\frac{\alpha\,V'}{f\,H^2} \sim\alpha\,\frac{M_P^2}{f^2}\gg 1\,,\nonumber\\
&&\nu_-\cong \frac{V''\,f}{\pi\,\alpha\,V'}\sim \frac{1}{\alpha}\ll 1\,,
\end{eqnarray}
where in the steps marked by the symbol $\sim$ we have used $V''\simeq V'/f\simeq V/f^2$.

The two point function of the metric perturbations has been computed, for our system, in~\cite{Anber:2009ua}. Here we will set up the calculations for the $n$-point function. See also~\cite{Barnaby:2011vw,Barnaby:2011pe} for a detailed analysis of this system in the regime where the backreaction of the gauge field on the inflaton is negligible.

\begin{figure}[ht]
\centerline{
\includegraphics[width=.5\textwidth]{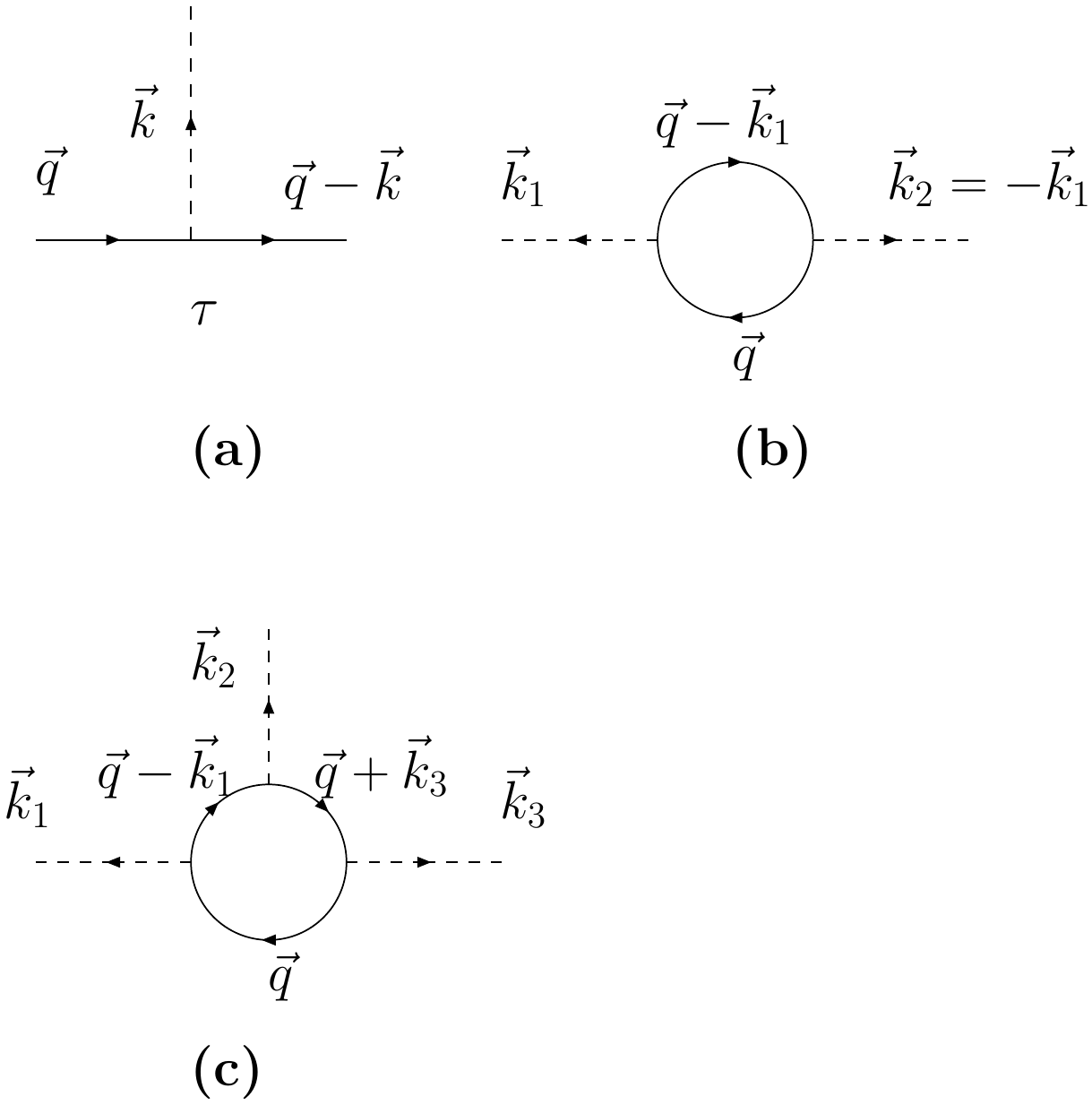}
}
\caption{Graphical representation of the correlation (\ref {grand correlation function}). The vertex (a) denotes the quantity ${\bf e}_+(\bf q)\cdot{\bf e}_+^*(\bf q-\bf k)\,{\cal I}(\tau,\,|\bf q|,\,|\bf q-\bf k|)$ as a function of the momenta $\bf q$ and  $\bf q-\bf k$. Diagrams (b), and (c) are the two-point and three-point functions. Dashed lines are the scalar perturbations $\phi(\bf k)$.}
\label{correlation diagrams}
\end{figure}

The $n$-point function of $\phi$ is given by 
\begin{eqnarray}
\langle \phi({\bf k}_1,\tau)\phi({\bf k}_2,\tau)&\dots&\phi({\bf k}_n,\tau)\rangle=\left(-\frac{\alpha}{f}\right)^n\int d\tau_1\,d\tau_2\,\dots\,d\tau_n\,a^2(\tau_1)\,a^2(\tau_2)\,\dots\,a^2(\tau_n)\\
\nonumber
&\times& G(\tau,\,\tau_1)\,G(\tau,\,\tau_2)\,\dots\,G(\tau,\,\tau_n)\,\langle J({\bf k}_1,\,\tau_1)\,J({\bf k}_2,\,\tau_2)\,\dots\,J({\bf k}_n,\,\tau_n) \rangle\,.
\end{eqnarray}
Generalizing the results of~\cite{Barnaby:2011vw} we obtain
\begin{eqnarray}
\nonumber
&&\langle \phi({\bf k}_1,\tau) \phi({\bf k}_2,\tau) ... \phi({\bf k}_n,\tau) \rangle={\cal N}\,{\sigma(n)}\,\left(-\frac{\alpha}{f}\right)^n \delta^{(3)}({\bf k}_1+{\bf k}_2+...+{\bf k}_n)\\
\nonumber
&&\quad\quad\quad\quad\quad\quad\quad\quad\quad\quad\quad\quad\times\int \frac{d^3 {\bf q}}{\left(2\pi\right)^{3n/2}}\left[{\bf e}_+ ({\bf q})\cdot {\bf e}^*_+({\bf q}-{\bf k}_1)\,{\cal I}(\tau,|{\bf q}|, |{\bf q}-{\bf k}_1|) \right]\\
\nonumber
&&\quad\quad\quad\quad\quad\quad\quad\quad\quad\quad\quad\quad\times\left[{\bf e}_+({\bf q}-{\bf k}_1)\cdot {\bf e}^*_+({\bf q}-{\bf k}_1-{\bf k}_2)\,{\cal I}(\tau,|{\bf q}-{\bf k}_1|, |{\bf q}-{\bf k}_1-{\bf k}_2|) \right]\\
\label{grand correlation function}
&&\quad\quad\quad\quad\quad\quad\quad\quad\quad\quad\quad\quad\times\dots\times \left[{\bf e}_+({\bf q}+{\bf k}_n)\cdot  {\bf e}_+^*({\bf q})\,{\cal I}(\tau,|{\bf q}+{\bf k}_n|, |{\bf q}|)  \right]\,,
\end{eqnarray}
where
\begin{eqnarray}
{\cal I}(\tau,\,k_1,\,k_2)=\int_{-\infty}^\tau d \tau' \frac{G(\tau,\,\tau')}{a^2(\tau')}\left\{k_1\,A_+(\tau',\,k_1)\,A_+'(\tau',\, k_2)+\left(k_1\leftrightarrow k_2\right)\right\}\,.
\end{eqnarray}
and where $\sigma(n)=(2(n-1))!!/2^n$ is a combinatorial factor. The  correlation function (\ref{grand correlation function}) can be represented graphically as in Figure \ref{correlation diagrams}.

Next, we compute ${\cal I}(\tau,\,k_1,\,k_2)$. Since we are interested in the spectrum at $k<<aH$, we neglect  the term $(\tau/\tau')^{\nu_+}$ in the expression~(\ref{nupm}) of the Green's function. Then using the expression~(\ref{approximate solution}) for $A_+$, we obtain
\begin{eqnarray}\label{cali}
{\cal I}(\tau,\, k_1,\,k_2 )\cong f\left[\sqrt{\frac{8\,\xi}{a(\tau)H}}\left( k_1^{1/2}+k_2^{1/2}\right)\right]\,\frac{H^2e^{2\pi\xi}}{2^{12}\nu_+\xi^4}\,\frac{k_1^{1/4}\,k_2^{1/4}}{\left(k_1^{1/2}+k_2^{1/2}\right)^7}\left\{|\tau|^{\nu_-}\left(k_1^{1/2}+k_2^{1/2}\right)^{2\,\nu_-} \right\}\,,
\end{eqnarray}
where
\begin{eqnarray}
f[x]=\int_x^\infty y^7 e^{-y}\,.
\end{eqnarray}
The quantity in curly brackets has a weak dependence on $\tau$, therefore we shall neglect its contribution to ${\cal I}$. In addition, the argument of $f[x]$ can be set to zero as we are interested in correlations as $\tau \rightarrow 0$. In this case we have $f[x]=7!$. We will denote by ${\cal I}(k_1,\,k_2)$ the function~(\ref{cali}) with $\nu_-=0$ and $f=7!$, and we will use this expression in eq.~(\ref{grand correlation function}) from now on.

\subsection{The two-point function}%

The two-point function is 
\begin{eqnarray}
\left\langle \phi({\bf k}_1)\,\phi({\bf k}_2) \right\rangle&=&{\cal N}\frac{\alpha^2}{2\,f^2}\, \delta^{(3)}({\bf k}_1+{\bf k}_2)\int \frac{d^3 {\bf q}}{\left(2\pi\right)^{3}}\left[{\bf e}_+ ({\bf q})\cdot  {\bf e}_+^*({\bf q}-{\bf k}_1)\,
{\cal I}(|{\bf q}|, |{\bf q}-{\bf k}_1|) \right]\nonumber\\
\label{two point function}
&&\times \left[{\bf e}_+ ({\bf q}-{\bf k}_1)\cdot  {\bf e}_+^*({\bf q})\,{\cal I}(|{\bf q}-{\bf k}_1|, |{\bf q}|)  \right]\,.
\end{eqnarray}
In order to compute the scalar products of polarization vectors appearing in this equation, we must know the general expression of ${\bf e}_+({\bf q})$. Writing the components of $\bf q$ as
\begin{eqnarray}
{\bf q}=q\left(\sin\theta_q\cos\phi_q,\sin\theta_q\sin\phi_q,\cos\theta_q\right)\,,
\end{eqnarray}
the polarization vector ${\bf e}_\pm({\bf q})$ reads
\begin{eqnarray}
{\bf e}_\pm({\bf q})=\frac{1}{\sqrt{2}}\left(\cos\theta_q\cos\phi_q\mp i\sin\phi_q,\cos\theta_q\sin\phi_q\pm i\cos\phi_q,-\sin\theta_q\right)\,.
\end{eqnarray}
Hence, we find
\begin{eqnarray}
\left|{\bf e}_+ ({\bf q})\cdot{\bf e}_+^*({\bf q}-{\bf k})  \right|^2=\frac{1}{4}\left[1+\frac{{\bf q}\cdot ({\bf q}-{\bf k})}{|{\bf q}||{\bf q}-{\bf k}|} \right]^2.
\end{eqnarray}
Computing numerically the angular integrals in eq.~(\ref{two point function}) we find
\begin{eqnarray}
\left\langle \phi({\bf k}_1)\,\phi({\bf k}_2) \right\rangle=2.1\times 10^{-6}\,{\cal N}\, \frac{\alpha^2}{f^2}\,\frac{H^4}{\nu_+^2}\,\frac{e^{4\pi \xi}}{\xi^8}\,\frac{\delta^{(3)}({\bf k}_1+{\bf k}_2)}{k_1^3}\,,
\end{eqnarray}
that matches the result obtained in~\cite{Anber:2009ua}.

\subsection{The three-point function}%

The three-point function is given by
\begin{eqnarray}
\nonumber
&&\left\langle \phi({\bf k}_1) \phi({\bf k}_2)\phi({\bf k}_3) \right\rangle=-\frac{\alpha^3}{f^3}\,\delta^{(3)}({\bf k}_1+{\bf k}_2+{\bf k}_3)\int \frac{d^3 {\bf q}}{\left(2\pi\right)^{9/2}}\left[{\bf e}_+ ({\bf q})\cdot  {\bf e}_+^*({\bf q}-{\bf k}_1)\,
{\cal I}(|{\bf q}|, |{\bf q}-{\bf k}_1|) \right]\\
&&\times \left[{\bf e}_+ ({\bf q}-{\bf k}_1)\cdot  {\bf e}_+^*({\bf q}+{\bf k}_3)\,{\cal I}(|{\bf q}-{\bf k}_1|, |{\bf q}+{\bf k}_3|)  \right]\left[{\bf e}_+({\bf q}+{\bf k}_3)\cdot  {\bf e}_+^*({\bf q})\,{\cal I}(|{\bf q}+{\bf k}_3|, |{\bf q}|)  \right].
\end{eqnarray}

The correlator depends on the size and shape of the triangle formed by the vectors ${\bf k}_1$, ${\bf k}_2$, and ${\bf k}_3$. We denote
\begin{eqnarray}
|{\bf k}_1|=k\,, \quad |{\bf k}_2|=x_2\,k\,,\quad |{\bf k}_3|=x_3\,k\,,
\end{eqnarray}
with
\begin{eqnarray}
{\bf k}_1&=&(k,\,0,\,0)\,,\qquad\qquad {\bf k}_2=-{\bf k}_1-{\bf k}_3\,,\\
\nonumber
{\bf k}_3&=&-\frac{k}{2x_3}\left(1-x_2^2+x_3^2,\,\sqrt{-(1-x_2+x_3)(1+x_2-x_3)(1-x_2-x_3)(1+x_2+x_3)},\,0\right)\,.
\end{eqnarray}

Using the equilateral configuration $x_2=x_3=1$, we find 
\begin{eqnarray}
\langle \phi({\bf k}_1)\,\phi({\bf k}_2)\,\phi({\bf k}_3) \rangle=-10^{-9}\,{\cal {N}}\,\frac{\alpha^3}{f^3}\,\frac{H^6}{\nu_+^3}\,\frac{e^{6\pi\xi}}{\xi^{12}}\frac{\delta^{(3)}({\bf k}_1+{\bf k}_2+{\bf k}_3)}{k^6}\,.
\end{eqnarray}

\section{Computation of the power spectrum and $f_{NL}$}%

Given the equations above it is straightforward to derive the expression for the power spectrum and the value of $f_{NL}$ for equilateral configurations in our system. The metric perturbation $\zeta({\bf k})$ is given by $\zeta({\bf k})=H\,\phi({\bf k})/\dot\Phi_0$. The scalar power spectrum can be read off from 
\begin{equation}
\langle \zeta({\bf k}_1)\,\zeta({\bf k}_1)\rangle={\cal P}_\zeta\,\frac{2\,\pi^2\,\delta^{(3)}({\bf k}_1+{\bf k}_2)}{k_1^3}\,,
\end{equation}
so that, using the background equation~(\ref{the main expression for xi}), the amplitude of the power spectrum turns out to be
\begin{equation}
{\cal P}_\zeta\simeq \frac{5\times 10^{-2}}{\xi^2\,{\cal N}}\,.
\end{equation}
Comparing this result to the observed value ${\cal P}_\zeta=2.5\times 10^{-9}$ we find that ${\cal N}\,\xi^2=2\times 10^7$. Given that $\xi$ cannot be larger than $20$ or so, we deduce that we need ${\cal N}\ga {\cal O}(10^5)$ to explain the smallness of the cosmological perturbations in this scenario.

Next, we can calculate $f_{NL}^{\mathrm {equil}}$. This quantity is related to the three point function of the curvature perturbation by
\begin{equation}
\langle \zeta({\bf k}_1)\,\zeta({\bf k}_2)\,\zeta({\bf k}_3)\rangle=\frac{3}{10}\,\left(2\pi\right)^{5/2}\,f_{NL}\,{\cal {P}}^2_\zeta\,\delta^{(3)}({\bf k}_1+{\bf k}_2+{\bf k}_3)\,\frac{\sum_i\,k_i^3}{\Pi_i\,k_i^3}\,\,,
\end{equation}
that, using the formulae above in the equilateral limit $k_1=k_2=k_3=k$, yields 
\begin{equation}
\langle \zeta({\bf k}_1)\,\zeta({\bf k}_2)\,\zeta({\bf k}_3)\rangle=-0.3\,\frac{1}{{\cal N}^2\,{\xi^3}}\,\frac{\delta^{(3)}({\bf k}_1+{\bf k}_2+{\bf k}_3)}{k^6}\,,
\end{equation}
and finally
\begin{equation}
f_{NL}^{\mathrm {equil}}= -1.3 \,\xi\,.
\end{equation}
Since, depending on the energy scale of inflation, $4\lesssim \xi\lesssim 20$, we have $5\lesssim |f_{NL}^{\mathrm {equil}}|\lesssim 30$. In particular, if $\xi\simeq 20$, as in the case of low scale inflation, then such a level of nongaussianities could be detectable by Planck.

The scaling of the $n-$point function can be justified by the following argument (see also~\cite{Barnaby:2011qe} for an analogous discussion of the two-point function in the case ${\cal N}=1$). Let us consider the main equation determining $\phi$, eq.~(\ref{largeeqphi}), in coordinate space and in cosmological time $t$ rather than in conformal time $\tau$. The most relevant terms are the friction term, on the left hand side, and the source $\delta_{\bf E\cdot\bf B}$ on the right hand side. Considering that perturbations evolve for about a Hubble time before freezing out of the horizon, we can approximate $\dot\phi\simeq H\,\phi$, so that we roughly obtain $\phi\simeq \delta_{\bf E\cdot\bf B}/V'$. When we compute the correlators, since the various components of $\delta_{\bf E\cdot\bf B}$ sum incoherently, we can estimate $\langle \phi^n\rangle\simeq \langle \delta_{\bf E\cdot\bf B}{}^n\rangle/V'{}^n\simeq {\cal N}\,\langle\delta_{\bf E\cdot\bf B}\rangle^n/V'{}^n$. Next we estimate $\langle\delta_{\bf E\cdot\bf B}\rangle\simeq \langle{\bf E}\cdot {\bf B}\rangle\simeq f\,V'/{\cal N}\alpha$, where we have used the background equation of motion given by the first of the eqs.~(\ref{equations of motion}). Finally, remembering that $\zeta=H\,\phi/\dot\Phi_0$, we obtain $\langle \zeta^n\rangle\simeq {\cal N}\,\left(\xi\,{\cal N}\right)^{-n}$, that matches the dependence on $\xi$ and ${\cal N}$ found above in the case of the two- and three-point function of $\zeta$. It is worth noting that, in the language of~\cite{Barnaby:2011pe}, the perturbations in this model are provided by a ``feeder'' mechanism that nonetheless leads to a hierarchical structure for the correlators.

\section{Gravitational waves}%

The generation of gravitational radiation in this system has been studied in~\cite{Sorbo:2011rz} in the regime where the backreaction of the gauge field on the inflaton was negligible. There it was found that the correlation functions of the left- and right-handed tensors take two contributions: on the top of the usual, parity symmetric part originating from the amplification of the quantum fluctuations of the graviton, there is a second, parity violating contribution generated by the presence of a classical gas of photons. The analysis of~\cite{Sorbo:2011rz} applies also to the present situation, and gives
\begin{eqnarray}
\nonumber
{\cal P}^{t,L}=\frac{H^2}{\pi^2 M_p^2}\left(1+8.6\times 10^{-7}{\cal N}\frac{H^2}{M_P^2}\frac{e^{4\pi\xi}}{\xi^6}\right)\,,\\
{\cal P}^{t,R}=\frac{H^2}{\pi^2 M_p^2}\left(1+1.8\times 10^{-9}{\cal N}\frac{H^2}{M_P^2}\frac{e^{4\pi\xi}}{\xi^6}\right)\,.
\end{eqnarray}
The tensor to scalar ratio $r=\left({\cal P}^{t,L}+{\cal P}^{t,R}\right)/{\cal P}_{\zeta}$ reads
\begin{eqnarray}
r=\frac{H^2}{\pi^2M_P^2}\frac{2+8.6\times 10^{-7}{\cal N}\frac{H^2}{M_P^2}\frac{e^{4\pi\xi}}{\xi^6}}{{\cal P}_{\zeta}}\,.
\end{eqnarray}
By trading $e^{4 \pi\xi}$ for the other quantities in the system using eq.~(\ref{the main expression for xi}), and remembering that ${\cal P}^{\zeta}\cong 0.05/({\cal N}\xi^2)$, we obtain 
\begin{equation}\label{tensors}
r=\frac{1}{{\cal P}_\zeta}\,\frac{2\,V}{3\,\pi^2\,M_P^4}+2.7\times 10^2\,\frac{\xi^4}{\alpha^2}\,\left(\frac{V'\,f}{V}\right)^2\,.
\end{equation}
The first term on the right hand side of the above equation corresponds to the standard spectrum of gravitational waves originating from the fact that the inflating Universe is a quasi de Sitter space. The second term corresponds to the contribution to the tensor spectrum induced by the presence of a gas of photons inside the horizon. Since $V'\,f\simeq V$, the current observational limit $r\la 0.2$ implies that $V^{1/4}\la 2\times 10^{16}$~GeV and $\alpha\ga 35\,\xi^2$.

In the specific cases discussed at the end of section 2.2, this implies that for high scale inflation $\Lambda\simeq 10^{16}$~GeV, $\xi\simeq 4$, we need $\alpha\ga 500$ (larger than the requirement on $\alpha$ originating from the need of realizing a sufficient number of efoldings of inflation). In the case of low scale inflation $\Lambda\simeq 10^{3}$~GeV, one has $\xi\simeq 20$ so that we need $\alpha \ga 10^4$. Again, this constraint on $\alpha$ is more stringent than that coming from the requirement of sufficient inflation.

Let us note that the second term on the right hand side of eq.~(\ref{tensors}) -- {\em i.e.}, the dominant contribution to $r$ when the gas of photons provides the main source of gravitons -- can take a wide range of values depending on $\alpha$ and $\xi$. This is different from the case with weak backreaction discussed in~\cite{Barnaby:2011vw,Sorbo:2011rz} where $r$, for large $\xi$, depends only on the slow-roll parameter. This discrepancy originates from the last term on the right hand side of eq.~(\ref{deltaeb}), that is negligible in the regime of weak backreaction, but becomes important in the situation studied in the present paper, and that affects the amplitude of the spectrum of scalar perturbations while leaving the tensor modes unchanged.

Remarkably, eq.~(\ref{tensors}) shows that primordial tensor modes might be detected in the upcoming surveys (that are expected to increase the sensitivity to $r$ by a factor of $10$ or so) even if the scale of inflation is very low, provided the ratio $\xi^2/\alpha$ is large enough. In this case the tensors would be essentially fully chiral -- a signature that  would be observable in future surveys once primordial tensors are detected~\cite{Sorbo:2011rz,Gluscevic:2010vv}.

\section{Conclusions and discussion}%

The model proposed in~\cite{Anber:2009ua} is able to produce inflation on a steep axion potential provided the coupling $\alpha$ is of order of a few hundred or larger. While in the case of a single family of $U(1)$ gauge fields the spectrum of scalar perturbations produced by this model is too large to account for observations, the presence of $\sim 10^5$ $U(1)$ gauge fields can reduce the amplitude of the perturbations to the observed value. 

One notable property of this model is that, while it depends strongly on the parameters $\alpha$ and ${\cal {N}}$, it has only a weak dependence on the axion constant $f$ and, more importantly, on the scale of inflation $\Lambda$. One might wonder whether the QCD axion could be a good inflaton candidate. We see two major obstructions against this idea. First, as we have seen, the reheating temperature $T_{\rm {RH}}$ is of the order of $\Lambda/10$, that for a QCD axion would give $T_{\rm {RH}}\sim 10$~MeV. While this is still large enough to allow for a successful Big Bang Nucleosynthesis~\cite{Kawasaki:1999na}, it is  hard to understand how to generate the observed baryon asymmetry in the short time between reheating at $\sim 10$~MeV and Nucleosynthesis at $\sim 1$~MeV. Of course, one might imagine that baryogenesis happens during the final stages of inflation, with the gas of chiral photons providing a strongly parity-violating environment, one of the necessary conditions for baryogenesis. The strongest obstruction, however, comes from the fact that we need $10^5$ copies of the $U(1)$ gauge field.  By the time Nucleosynthesis occurs, all of these fields (with the exception of the one associated to the Standard Model photon) must have decayed into ordinary matter. While this is definitely possible (if the gauge fields have a small mass, they can decay into matter) the short time window between $T\sim 10$~MeV and $T\sim 1$~MeV is probably too narrow to allow for this process to happen completely.

In the present work we have studied the nongaussianities and the tensor modes generated in this scenario. The phenomenology of the scenario turns out to be rather rich: the (equilateral) bispectrum is consistent with the current limits, but its amplitude is sufficiently large to be detectable by Planck. In fact, the scenario described in this paper would be under severe pressure if future data were to constrain $|f_{NL}^{\rm equil}|$ to be smaller than $5$ or so. The non-observation of tensors imposes constraints on the parameter $\alpha$ that are about one order of magnitude stronger than those found in~\cite{Anber:2009ua}. The detectability of tensors is strongly parameter-dependent. Remarkably, however, tensors might be large enough to be detectable (say with $r\ga 0.01$ or so) even if inflation occurred at very low energies (similarly to other situations considered in~\cite{Senatore:2011sp}). As pointed out in~\cite{Cook:2011hg}, this originates from the fact that, due to the conservation of angular momentum, vectors are (unlike scalars) an efficient source of tensors. More importantly, tensors produced in this scenario would be almost fully chiral -- a property that might be detected in future surveys~\cite{Gluscevic:2010vv} and that would provide a very specific signature of this model, especially if chiral tensors were to be detected along with some degree of nongaussianity in the scalar spectrum.

\smallskip

{\bf Acknowledgments.} It is a pleasure to thank Neil Barnaby and Enrico Pajer for useful discussions. The work of M.A. is supported by NSERC Discovery Grant of Canada. The work of L.S. is partially supported by the U.S. National Science Foundation grant PHY-0555304.


\end{document}